\documentclass[12pt]{article}
\usepackage[margin=1in]{geometry}                 
\usepackage{setspace}
\usepackage{graphicx}
\usepackage{amsmath}
\usepackage{amssymb}
\usepackage{bm}
\usepackage{url}
\usepackage{epstopdf}
\usepackage{times}
\usepackage{courier}
\usepackage{authblk}

\usepackage{algorithmic}
\usepackage{textcomp}
\usepackage{xcolor}
\usepackage{subcaption}
\usepackage{diagbox}
\usepackage{tabulary}
\usepackage{mathtools}
\usepackage{listings, multicol}
\title{SkyMemory: A LEO Edge Cache for Transformer Inference Optimization and Scale Out}
\author{Thomas Sandholm, Sayandev Mukherjee, Lin Cheng, Bernardo A. Huberman}
\affil{}

\begin{document}
\maketitle

\begin{abstract}
We expand the scope of cache memory to include LEO constellations, which are highly distributed systems with thousands of satellites connected with free-space optics inter-satellite links (ISL) always only one hop from any point on earth. We show how to increase the number of cache hits and improve the speed of inference for the important use case of LLMs. These benefits apply not only to LLMs, both terrestrially hosted and on satellites, but also generalize to any cache distributed over multiple locations that needs to be accessed in a timely manner. We show the benefit of our key value cache (KVC) protocol in simulations and present a proof-of-concept implementation of the protocol for KVCs on a testbed comprising 5 Intel NUC Linux mini PCs hosting a 19x5 constellation, with an NVIDIA Jetson Nano 8GB GPU hosting the LLM.
\end{abstract}

\section{Introduction}
Models based on the Transformer architecture~\cite{vaswani2017} have become the de facto standard for AI workloads. 
The versatility of the Transformer can be explained by an attention mechanism tracking
correlations between tokens at different positions in a sequence. Originally conceived
with text tokens for language translation, Large Language Models (LLMs) using the Transformer
architecture are now adapted to solve almost any natural language processing (NLP) tasks, and
have now been generalized to be Large Multimodal Models (LMMs) that process text and images,
both still and video.  We will discuss the text LLMs here, but the proposed protocol is applicable
to any model employing the Transformer architecture, or any application that can benefit from
edge caching.

An LLM processes a text prompt by first converting it to a sequence of tokens representing the
semantics of the text. Next comes the prefilling phase, where this token sequence is passed
through the attention layers of the LLM. The prefilling phase results in an output token
sequence that is generated one token at a time (next token prediction). Converting the
generated tokens to text again, and controlling which token to produce given probability
distributions over tokens is referred to as the decoding phase. The whole process of generating an
output text from a prompt text is referred to as inference.
With large prompts that are repetitive across generations, e.g. containing entire documents
like in retrieval augmented generation (RAG) or long chat histories (contexts),the conversion
of a token sequence to an output suffers large overheads. 

In general, the attention layers cause the computational overhead to increase
quadratically with the length of the context. As a result, early work to optimize this step
cached the LLM intermediate outputs for common contexts in what is called key value caches (KVC),
popularized in a seminal work called PagedAttention~\cite{kwon2023}.

KVC stores KVC for fixed blocks of context similarly to how an operating system uses page
caching. An important feature of this cache is that the sequence of blocks matters. 
So, a prompt is converted to an ordered sequence of blocks. When a new prompt appears it is
enough to find a matching block in the cache that is furthest toward the end of the sequence of
blocks, because that means that all preceding blocks up to and including the matching block are
also in the cache. 

Even for very small models (1B) the KVC for a small block (128 tokens) can be
rather large (~3MB). The cache retrieval must be fast (from memory) so as not to defeat the
purpose of speeding up the GPU computation. A trend to scale the inference to large models and
contexts is to store the KVC across a memory hierarchy, such as GPU, CPU, network storage, cold
disk storage, and move the caches between the layers based on various last recently used (LRU)
eviction policies. For large contexts, the usual arguments for speed and computational complexity
apply when comparing computation from scratch (quadratic complexity) versus retrieving from a
cache, which is typically a lookup from a hash table (O(1) complexity).Thus, by trading memory
storage for computation, caching can improve the latency and computational complexity of inference
by an order of magnitude.

Recent advancement of low-earth-orbit (LEO) constellations for communication~\cite{vanelli2024} and earth
observation has given rise to the concept of LEO edge, which makes use of the vast computational
resources circling in orbit across the earth. Even full orbit-bound data centers have been
proposed\footnote{\url{https://www.starcloud.com/}}. The advantage of a LEO constellation is
that it is a highly distributed system with thousands of satellites connected with free-space
optics inter satellite links (ISL)~\cite{perdigues2021} and always a hop a way from any point on earth. Highly
directional phased array antennas have pushed down the latency to single digit milliseconds for
ground communication. The key challenge is handling satellite mobility, as a particular LEO satellite may only be visible
from a point on earth for 5-10 minutes.

In this work, we expand the scope of cache memory to include the memory on LEO satellites.
This increases cache hits, which in turn improves speed of inference for LLMs, in particular
the time to first token (TTFT) in the prefilling stage.  This benefit applies to LLMs hosted both
terrestrially and on satellites, and it is generalizable not just to key value caches for LLMs
but to any use case with a cache distributed over multiple locations that needs to be accessed
and set quickly.

\section{Motivation}
A key-value cache can be stored in memory hierarchies and our 
solution can be integrated into a stack of both faster and slower memory.

Table~\ref{T:memorystack} summarizes an approximate latency mapping of different 
memory types, including the LEO edge that is the focus of this work.

\begin{table}[htbp]
        \caption{Approximate latency for different memory types.}
\begin{center}
\begin{tabular}{|l|l|}
\hline
  Type & Latency \\
\hline
  CPU & 10-15 nanoseconds \\
  GPU & 50-100 nanoseconds \\
  RDMA & 2-5 microseconds \\
  SSD & 20-200 microseconds \\
  HDD & 2-20 milliseconds \\
  NAS & 30-40 milliseconds \\
  LEO (current RF) & 20-50 milliseconds \\
  LEO (theoretical Laser) & 2-4 milliseconds \\
\hline
\end{tabular}
\label{T:memorystack}
\end{center}
\end{table}

From a single point on Earth, as many as 10-20 LEO satellites may be visible,
which would allow direct communication with multiple satellites at a time.

The LEO numbers in Table~\ref{T:memorystack} are for ground-to-satellite communication. 
In the on-board LLM case, for FSO in the LEO constellation ISL, the latency is determined by inter-orbital 
and intra-orbital distances and speed of light. 
The inter-satellite distances can be computed with the following equation from~\cite{pfandzelter2022}: 
\begin{equation}\label{eq:dm}
	D_m = (r_E + h)\sqrt{2\left[1-\cos\left(\frac{2\pi}{M}\right)\right]}
\end{equation}
where $D_m$ is the distance between satellites within the same plane, $r_E$ the radius of earth,
$h$ the altitude of the constellation, and $M$ the number of satellites within the same plane in
the constellation.  The distance between two neighboring satellites within different planes
varies during an orbital cycle but the maximum distance is given by
\begin{equation}
	D_n = (r_E + h)\sqrt{2\left[1-\cos\left(\frac{2\pi}{N}\right)\right]}
\end{equation}
where $N$ is the number of planes in a constellation. Hence, we can consider~\eqref{eq:dm}
as a worst-case scenario distance or latency for all ISL communication.

We also adopt the +GRID 2D-Torus Mesh network architecture described in~\cite{pfandzelter2022},
which assumes that each satellite has four interconnects used to communicate (through laser) with
its closest neighbors.

\begin{figure}[htbp]
        \centerline{\includegraphics[scale=0.75]{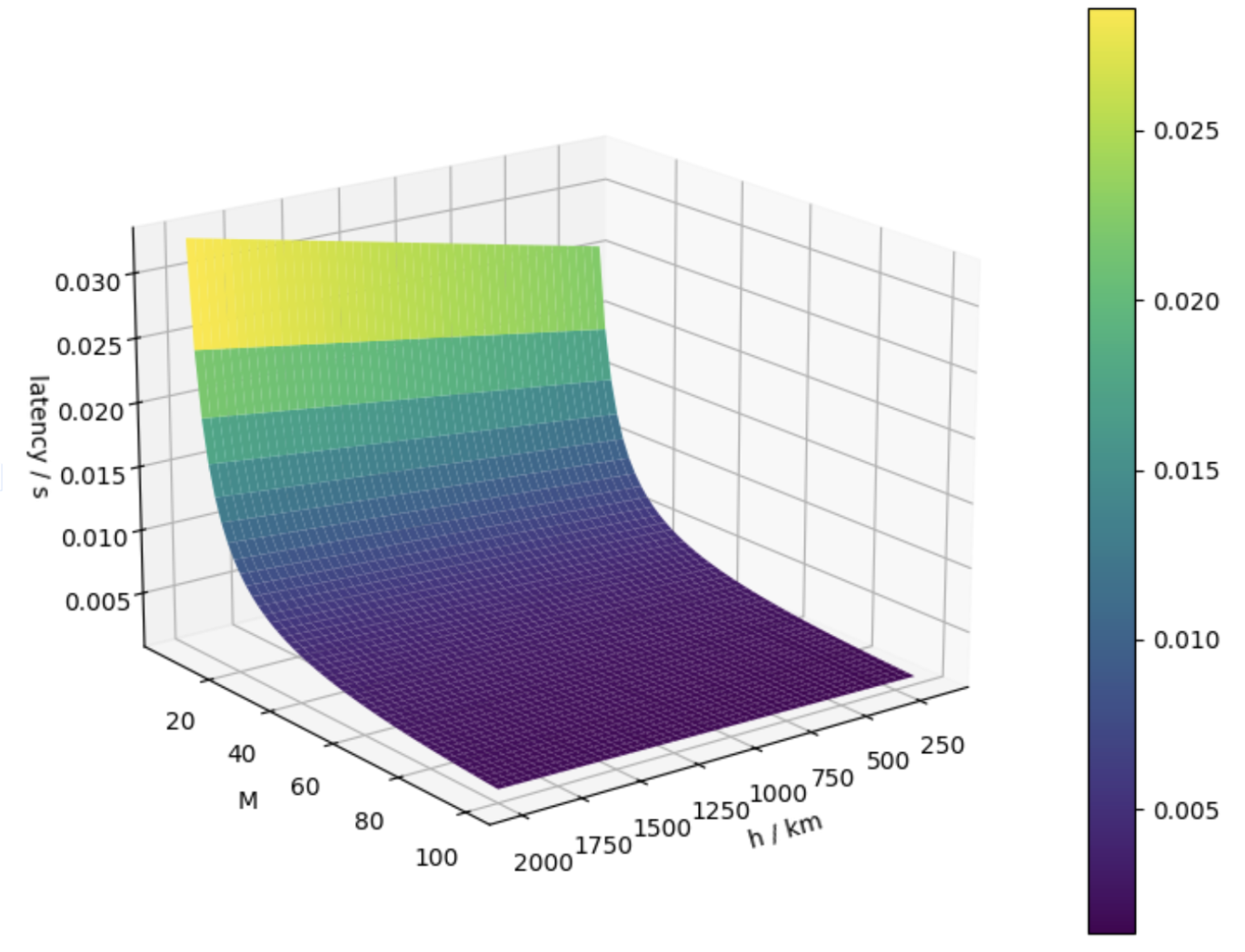}}
        \caption{Dependence of the intra-plane latency on $M$ and $h$ in a three-dimensional plot
	of latency vs constellation altitude $h$ for different fixed values of $M$, the number of satellites in a single plane of the LEO constellation.}
\label{latency1}
\end{figure}

\begin{figure}[htbp]
        \centerline{\includegraphics[scale=0.75]{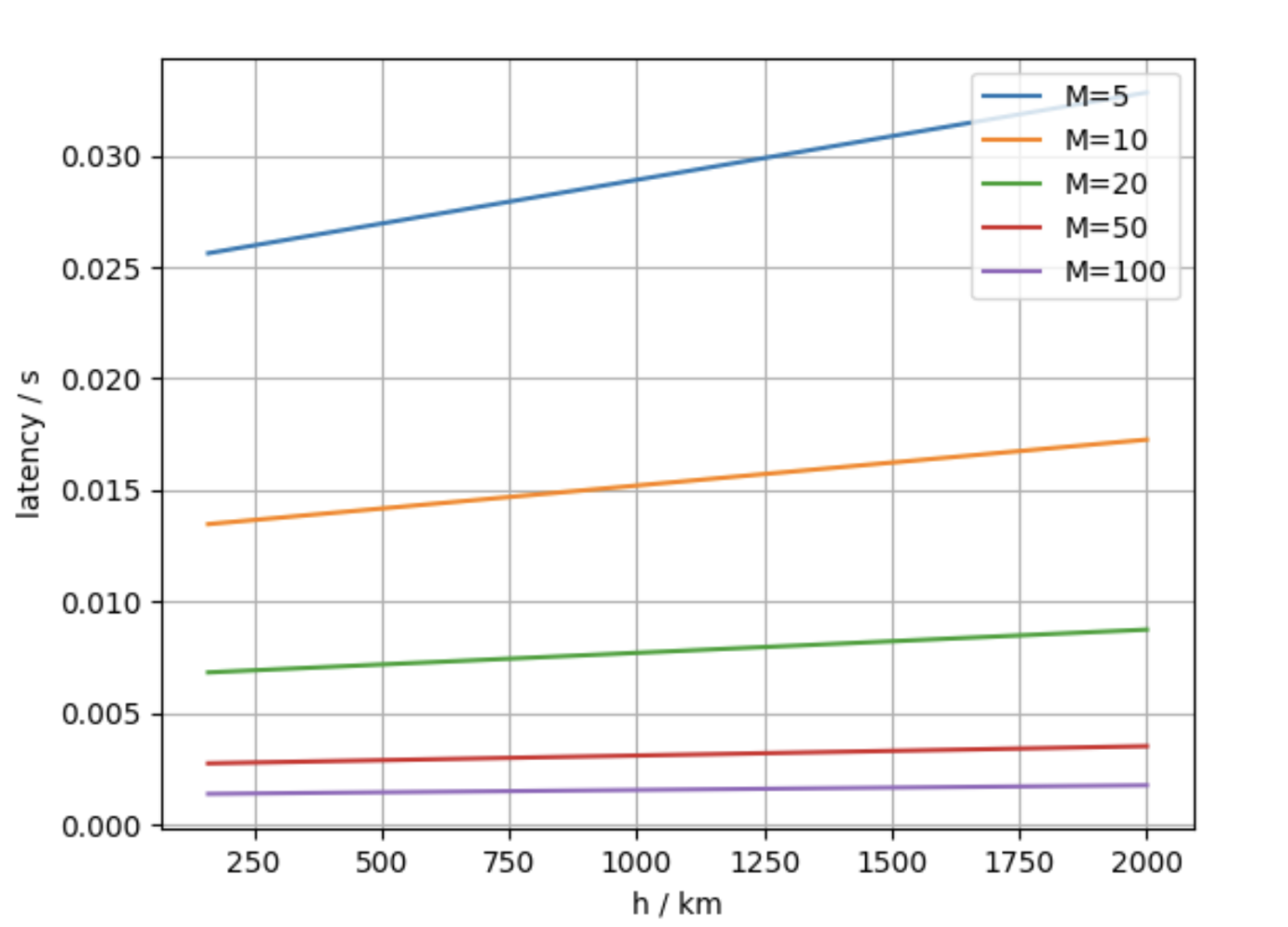}}
        \caption{Dependence of the intra-plane latency on $M$ and $h$ in a contour plot
	of latency vs constellation altitude $h$ for different fixed values of $M$, the number of satellites in a single plane of the LEO constellation.}
\label{latency2}
\end{figure}

Figures~\ref{latency1} and \ref{latency2} illustrate the worst-case ISL latency as a function of $M$ and $h$.


Extrapolating the numbers from these simulations, we can see that
we get roughly a latency between SSD and HDD (see Table~\ref{T:memorystack})
with about $50+$ satellites in a plane or $50+$ planes ($<2$ milliseconds).

\section{Design of SkyMemory, a LEO hosted KVC}
We propose a LEO satellite constellation hosted KVC that can fit seamlessly into
current KVC memory hierarchies, for optimized LLM inference in general and prefilling
in particular. Rather than hosting the KV Lists of an LLM during token generation,
we find, pull in, and deploy (in the GPU or other computing hardware when the
inference computation takes place) the most appropriate full KVC of all KV Lists
given a prompt, before running the inference computation on the prompt.
We present a protocol for two use cases, LLMs hosted on earth and LLMs hosted on board satellites in a constellation.

Note that this protocol is applicable to reduce the latency of any cache
(not just a KVC for LLMs) that is distributed over different physical
locations by exploiting the very low latencies of LEO links (see Table~\ref{T:memorystack}).

\subsection{Baseline Protocol}
Prompts are split into token blocks of a fixed size (e.g., 128 tokens). 
We start by hashing the first token block, then for the second token
block, we hash that token block together with the hash of the first token
block.  Thus, the hash for the second token block is actually the hash of the
first two token blocks.  Similarly, we hash the third token block together
with the hash for the second token block (which is actually the hash of
the first two token blocks), and so on, such that the hash for a token block
in the cache is actually a hash of all the token blocks up to and including
that token block.  When finding a matching KVC it is thus enough to find
the matching hash that is furthest to the end of the hash list. The lookup
input is an ordered list of hashes representing the input prompt.

Given that these blocks can be large (several MB or GB even for modestly sized LLMs), we
further split up each block into chunks of fixed byte size (e.g. 6k bytes). Mapping from a
chunk to a target server and how to look up a chunk can be done differently depending on
the use case, discussed below. However, it is worth noting that a failed lookup of a single
chunk is enough to determine that the KVC does not exist for the queried block.

A baseline implementation of the protocol just computes the server (virtual chunk
destination that will be mapped to a physical destination such as a satellite) to store on
as chunk\_id mod $n$ where $n$ is the maximum value of the index of the server to host the chunk. Note that this
allows for parallelism both in setting and getting a single KVC.
Each KVC cache entry is hence identified by the tuple (block\_hash, chunk\_id).

This baseline protocol was inspired by the vLLM implementation of a prefix
caching block~\cite{kwon2023}.

\subsection{LEO Constellation Model}
A LEO constellation comprises a set of satellites orbiting earth in near circular paths.
A constellation is identified by an altitude and an inclination angle shared across all its
orbital paths. Each orbital path hosts the same number of equidistant satellites in a grid
formation. There is wraparound both within a plane (first and last satellites can communicate)
and across planes (first and last planes can communicate) to form a 2D-Torus Mesh often
referred to as +GRID. Even though ISL communication is performed via free space optics,
geography matters and the closer the satellite is that hosts the cache to the one requesting
it the better and more reliable the latency will be.

In general, we cannot even assume that the whole cache of chunks for a block resides within
a single satellite. Distributing the cache saves memory but also improves parallelism in
setting and getting values. A cache miss is not catastrophic as a KV cache can always be
recomputed. So, while redundancy is not required for reliability, it can  improve latency.
The more blocks you can cache, the higher likelihood of cache hits, and therefore the lower
the inference time.

\begin{figure}[htbp]
	\centering
        \includegraphics[scale=0.65]{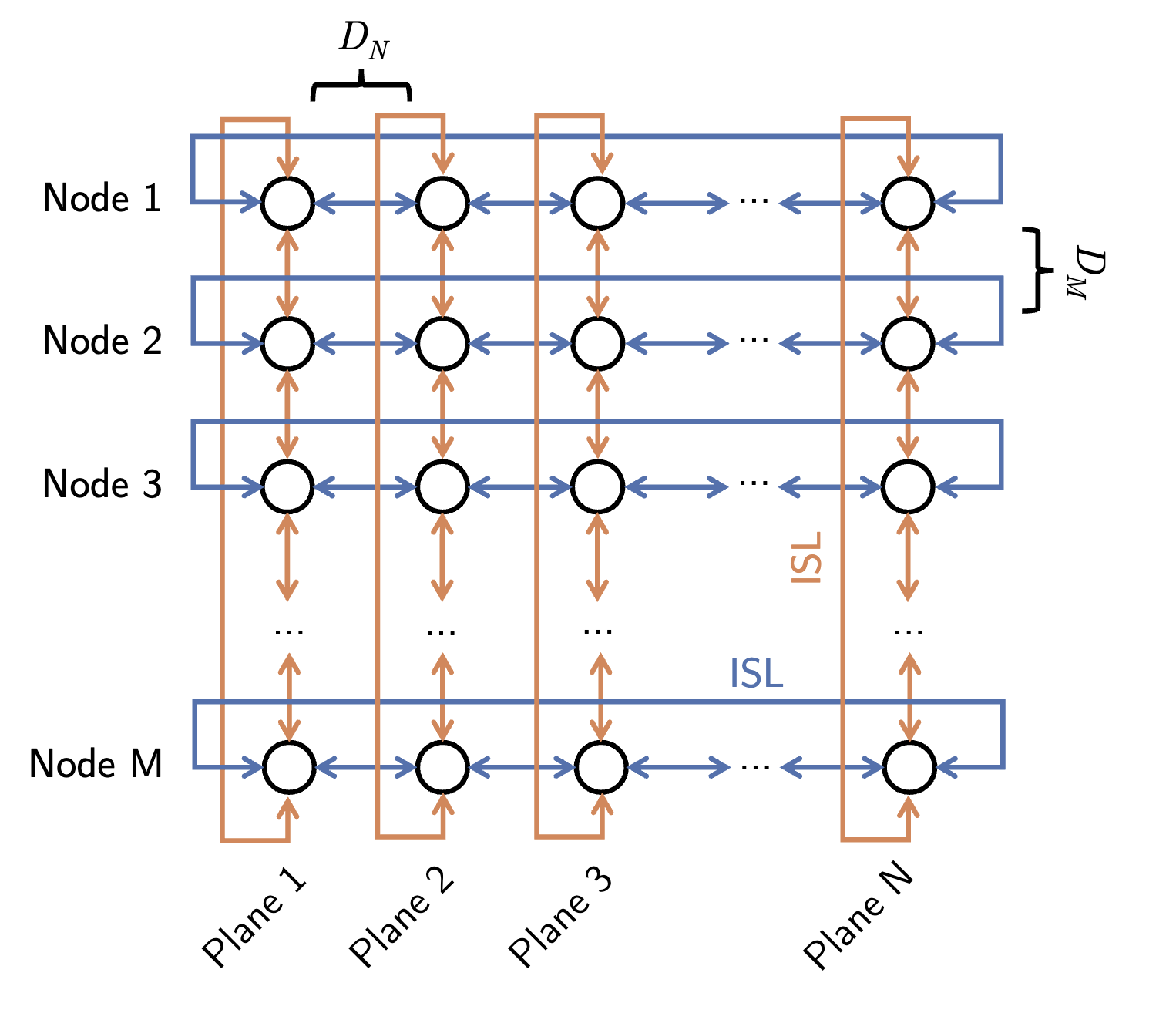}
        \caption{The +GRID networking model, so named because the 4 ISL links from each satellite look like a plus sign.  Note that the topology is a torus, so the topmost and bottom-most rows wrap around, as do the leftmost and rightmost columns. Source: Pfandzelter and Bermbach~\cite{pfandzelter2022}.}
\label{torus}
\end{figure}

The networking model of a 2D-Torus mesh with 4 ISL links from each satellite (+GRID) is illustrated in Figure~\ref{torus} below (from~\cite{pfandzelter2022})

Although the ideal architecture would be to connect and communicate directly with all line-of-sight satellites from the ground directly, the protocol is structured such that all the cache endpoints are within the fewest possible routing hops from the closest satellite, which could be used if the line of sight (LOS) is obstructed.

\subsection{Interface}
The abstract interface provides two operations to obtain a KVC for a prompt (or empty KVC
if none was found) and to add blocks in the cache based on a prompt.
Both operations operate on a given model and tokenizer. If any parameter changes in the
model, the cache is no longer valid. Similarly, a different tokenizer would also invalidate
the cache.

\begin{verbatim}
class KVCManager:
    init(model:LLMModel, tokenizer:LLMTokenizer)
    add_blocks(prompt:String)
    get_cache(prompt:String) ->KVC
\end{verbatim}

The KVC returned can be passed into LLMModel:generate calls to speed up generations, often referred to as past key values. The KVC can be implemented to be memory efficient by trading off accuracy using various quantization techniques. 

\subsection{Chunk to Server Mapping and Migration}
Servers are virtual satellite destinations, and the number of servers determines how the
chunks are spread out in a deployment. The initial chunk is always stored in the current
closest LOS satellite. The server ids are then distributed around this point in concentric
circles to lay out the chunks.

Once a new set of satellites is in LOS the chunks need to be migrated for efficient retrieval.This can be done in parallel in each orbital plane.

Next, we will describe the foundation of our contribution: how chunks are mapped to satellites
for different use cases. We provide three different mappings, namely {\it rotation-aware},
{\it hop-aware}, and {\it rotation- and hop-aware caching}.

\subsection{Rotation Aware Caching}
\begin{figure}[htbp]
	\centering
        \includegraphics[scale=0.5]{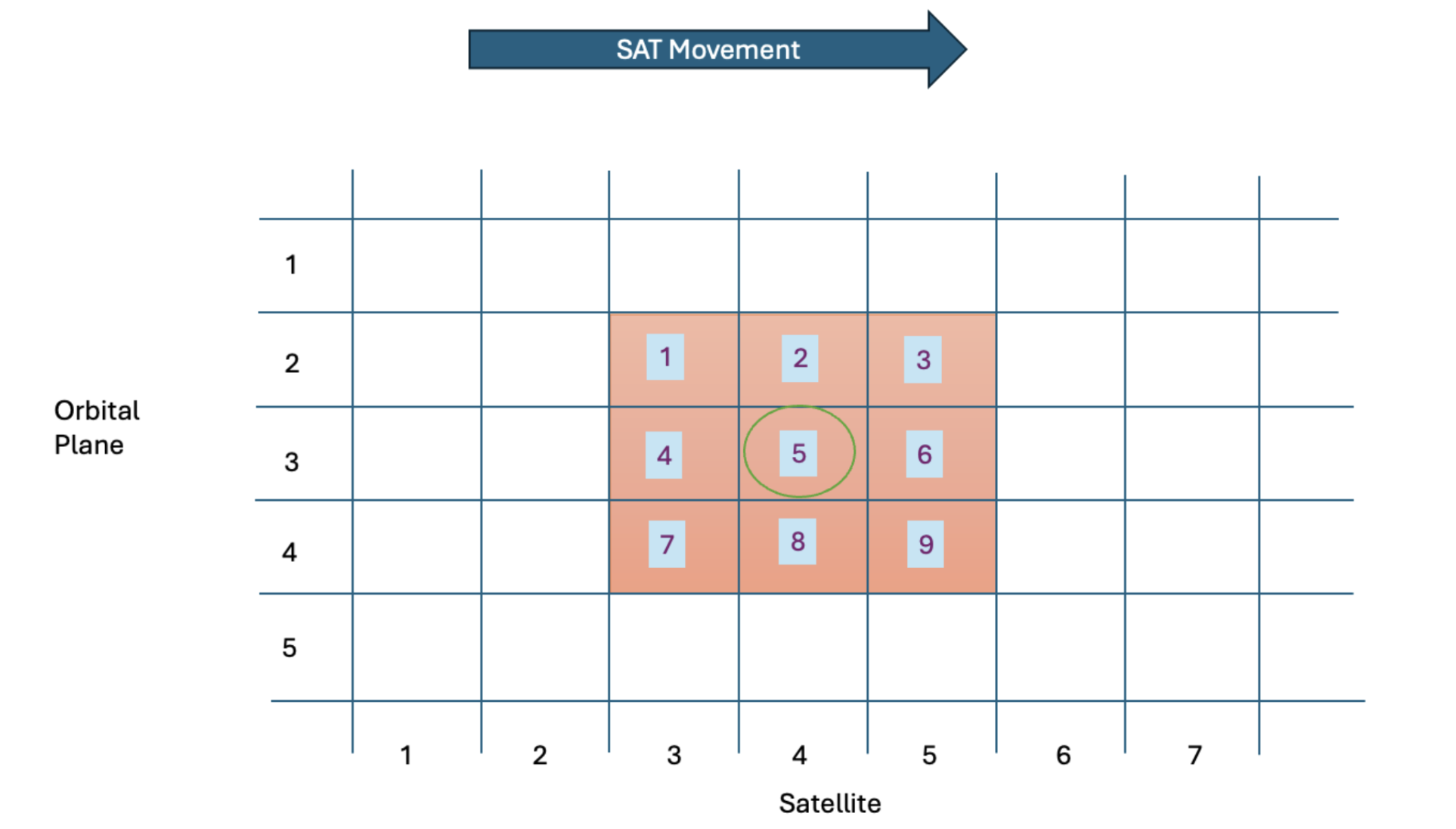}
        \caption{Each square of the grid represents one satellite.  The squares colored orange are in the LOS from the ground-hosted LLM, while the square representing the satellite closest to the LLM is circled in green. The numbers in the squares represent the logical chunk indices on the servers in the satellites.}
\label{rotationbefore}
\end{figure}

In the rotation-aware caching mapping, servers (logical chunk indexes)
are mapped to satellites left to right, top to bottom within a LOS grid,
see Figure~\ref{rotationbefore}. In this and subsequent figures, the circled satellite in the center is the one that is closest to the LLM host on the ground, i.e., from the perspective of the ground-hosted LLM, the circled satellite may be assumed to be directly overhead.

\begin{figure}[htbp]
	\centering
        \includegraphics[scale=0.5]{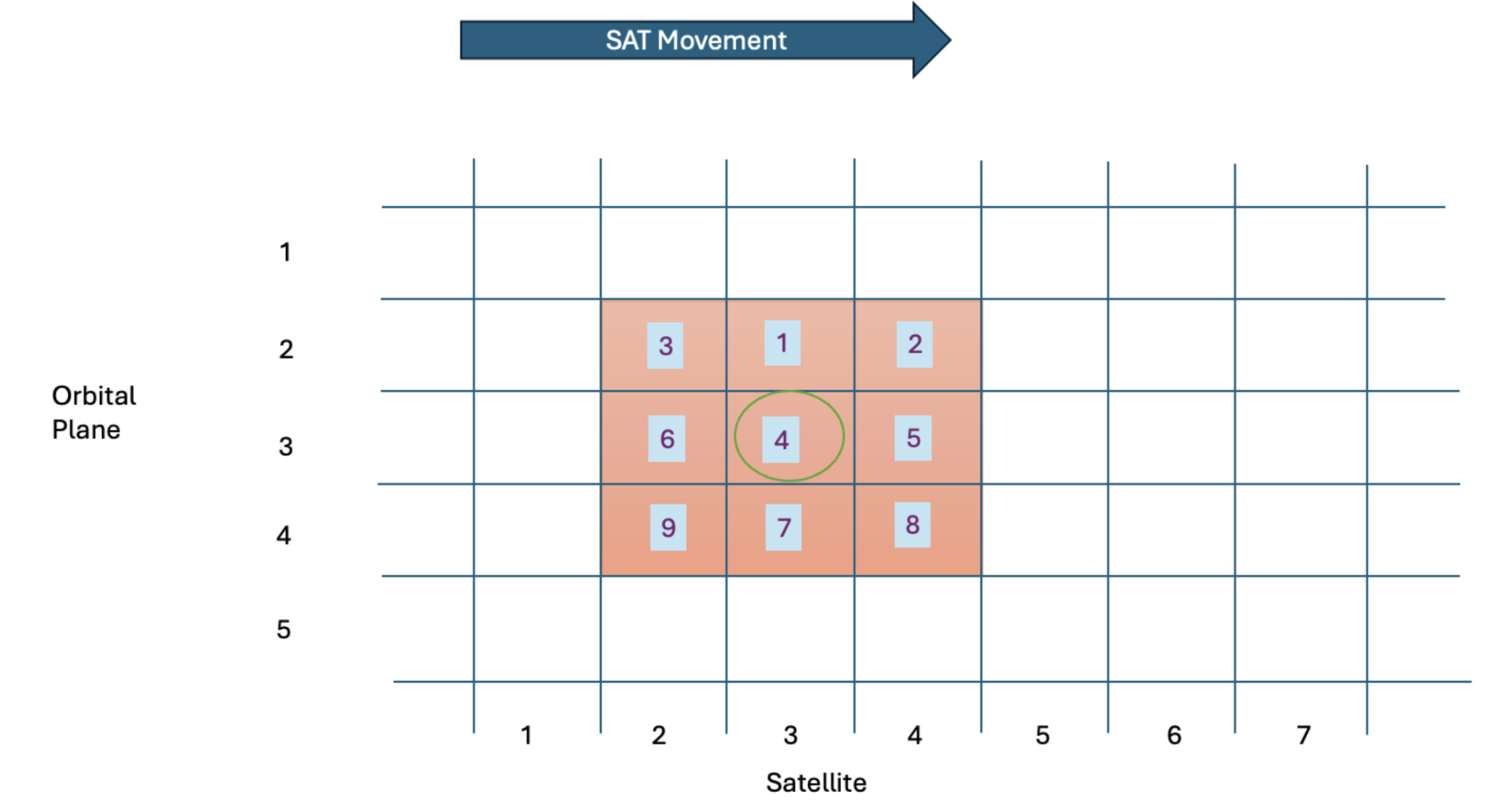}
        \caption{Illustrating chunk migration from the rightmost of the colored columns to the leftmost colored column, to account for rotation of the satellites in Figure~\ref{rotationbefore}.}
\label{rotationafter}
\end{figure}

This mapping works best in the use case of a ground hosted LLM that has LOS to a large
number of satellites reliably (e.g. 10-20). Chunk migration is done from the column
furthest to the right to the column furthest to the left. See Figure~\ref{rotationafter}. 

\subsection{Hop Aware Caching}
\begin{figure}[htbp]
	\centering
        \includegraphics[scale=0.5]{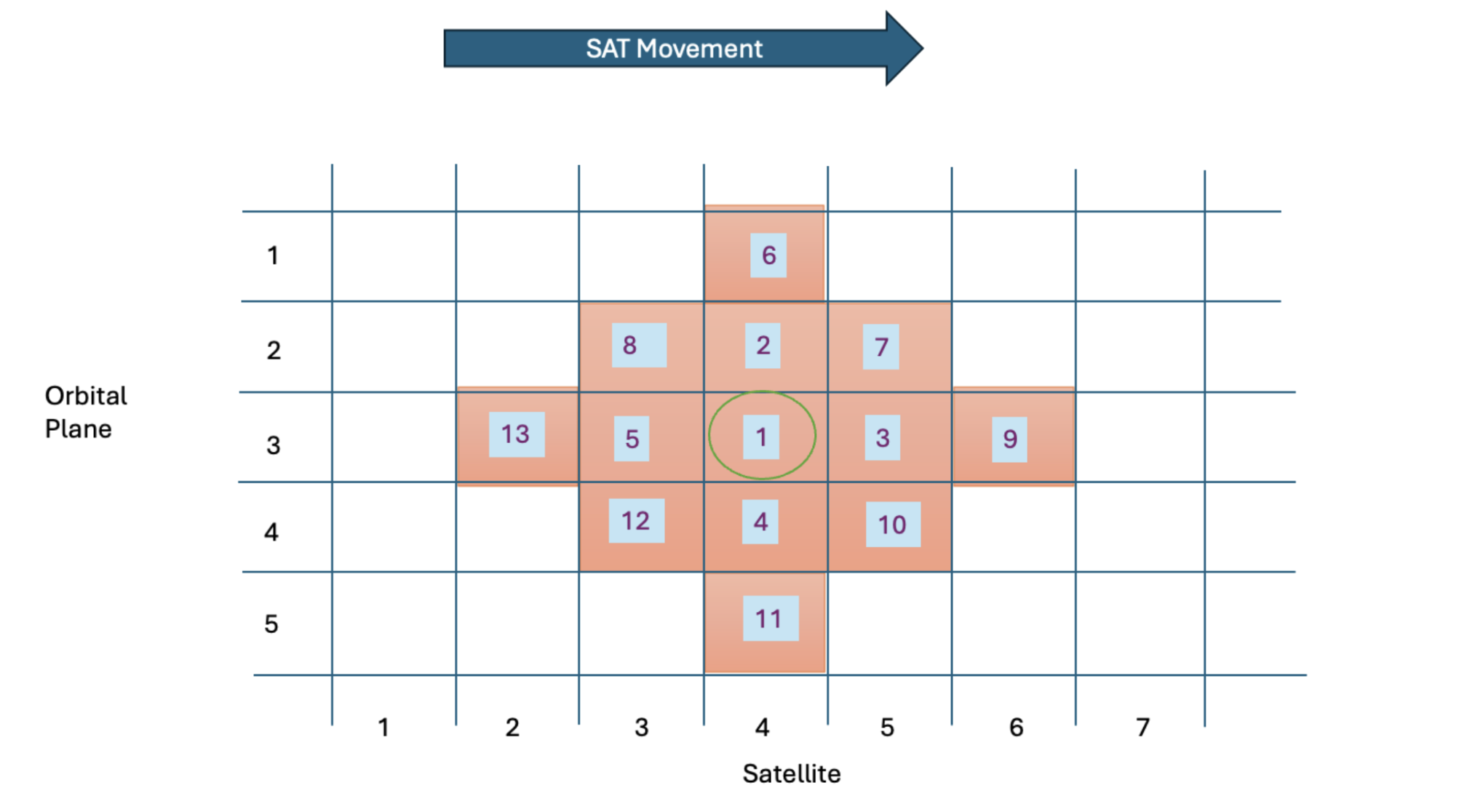}
        \caption{Labeling of concentric circles of servers around the satellite in the center.}
\label{hop}
\end{figure}

In the hop-aware caching mapping servers are mapped to satellites in concentric
circles starting from a given satellite, as shown in Figure~\ref{hop}.
This mapping works best when no migration is necessary and the LLM is hosted on-board a
fixed satellite.

Note that concentric circles and hops may be logical, so that faster horizontal within-plane hops can result
in wider horizontal areas. 
 
\subsection{Rotation and Hop Aware Caching}
In rotation- and hop-aware caching, servers are mapped to satellites in concentric
circles starting from a given satellite and within a LOS bounding box. The bounding box is determined by taking
the square root of the total number of servers and centering the box around the satellite closest to the (ground-hosted) LLM location. 

\begin{figure}[htbp]
	\centering
        \includegraphics[scale=0.5]{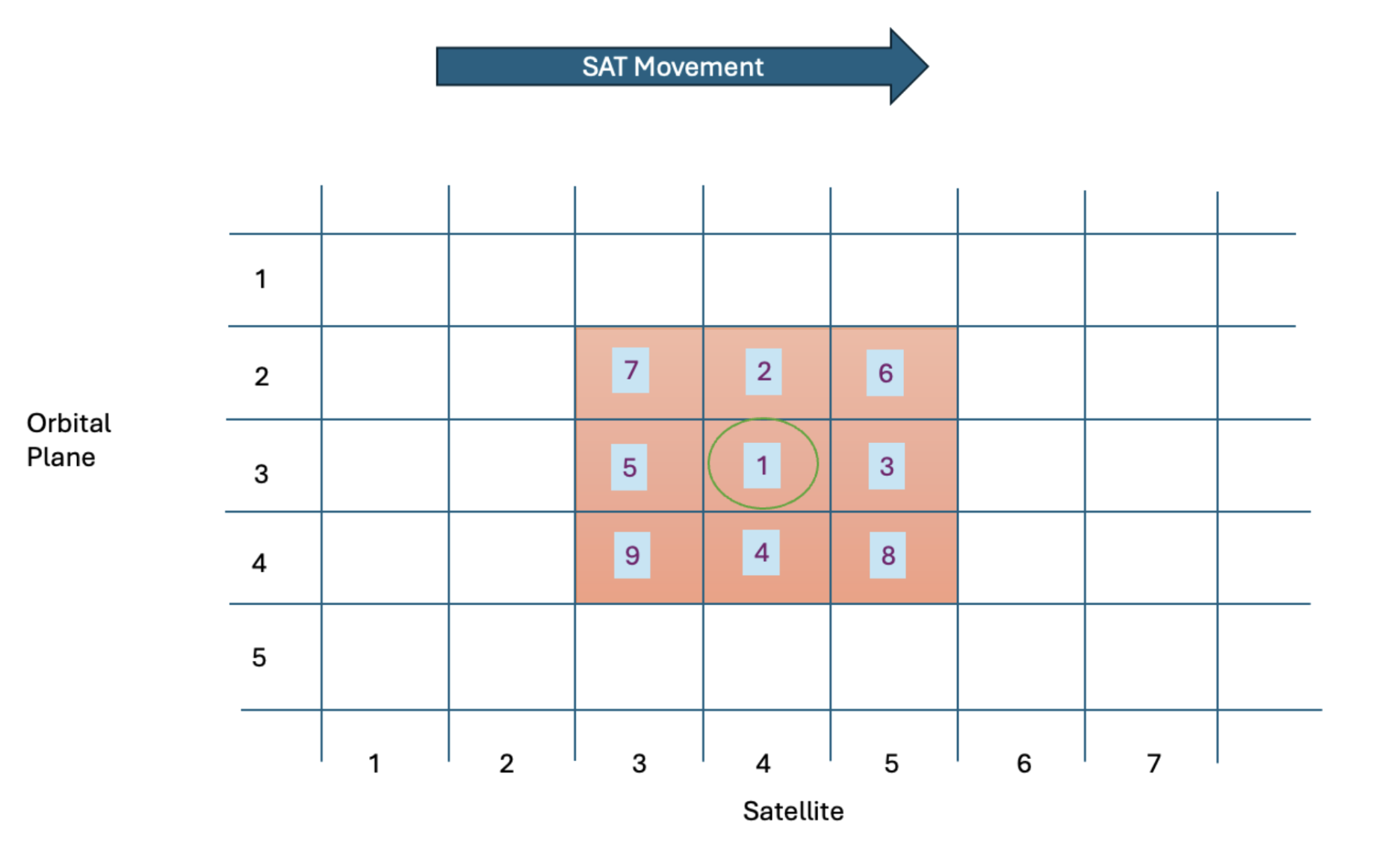}
        \caption{Rotation and Hop Aware: Initially, the green satellite (satellite 4 in orbital plane 3) is directly overhead.}
\label{hoprotationbefore}
\end{figure}

\begin{figure}[htbp]
	\centering
        \includegraphics[scale=0.5]{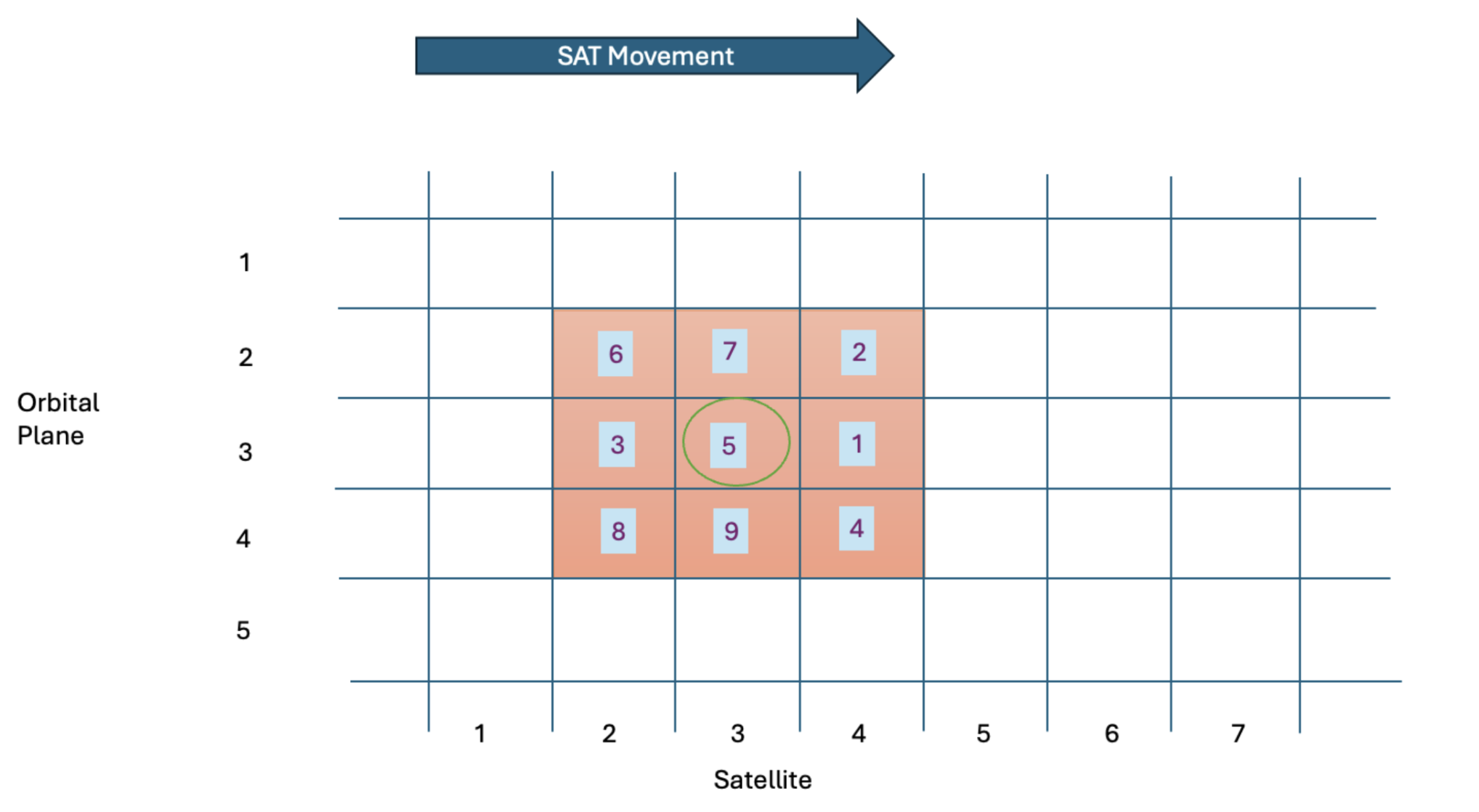}
        \caption{Rotation and Hop Aware: Given the direction of satellite movement, the column of satellites to its left (i.e., satellite 3 in the orbital plane 3) will be directly overhead in a few minutes. Thus to prevent chunks 6, 3, and 8 from going out of LOS, these chunks are migrated to satellite 2 in orbital planes 2, 3, and 4.}
\label{hoprotationafter}
\end{figure}

This works best in the ground-to-satellite scenario like the rotation case above, but where the
satellites cannot be reached reliably with a single hop from the ground, see Figures~\ref{hoprotationbefore} and~\ref{hoprotationafter}.

In Figure~\ref{hoprotationafter} we note that, as the satellites furthest to the east move out of sight their
chunks are migrated to the satellites about to enter LOS on the west. In this case Satellite (sat 5,orb 2) migrates 
chunk 6 to (2,2), 3:(5,3)$\rightarrow$(2,3) and 8:(5,4)$\rightarrow$(2,4). Note that all these migrations can be done in parallel and 
there is no harm in the chunk being stored in two satellites for some period of time.

Lastly, we note that if we predict a cache hit on a certain set of chunks at some future time 
(for example, because of a predictive algorithm), then we can exploit the fact that the set of satellites
in the LOS at that future time is known exactly and arrange to make those chunks available on those LOS
satellites at that time. 

\subsection{Protocol}
\begin{figure}[htbp]
	\centering
        \includegraphics[scale=0.5]{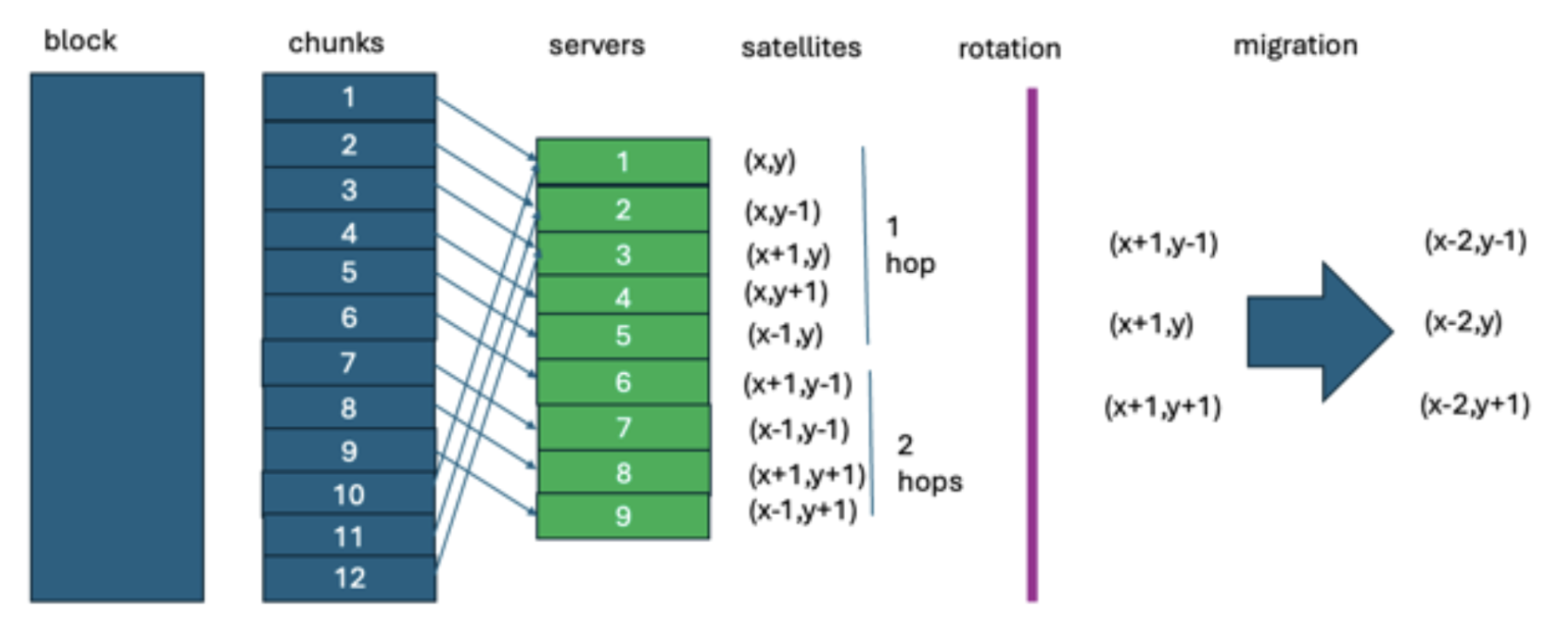}
        \caption{Setting Block + 1 Rotation Migration: How chunks are created from blocks, mapped to logical server and then satellites, and how the new satellite hosts for a chunk can be computed after a rotation.}
\label{setblock}
\end{figure}

The protocol for setting and getting KVC content can be summarized as follows:

\subsubsection*{Set KVC (Figure~\ref{setblock}):}
\begin{enumerate}
\item{The prompt is tokenized and split into equal-sized token blocks}
\item{Each block is ordered and hashed based on the previous block hash and the token list in the current block. The initial block has a null hash 0 as the previous block hash.}
\item{A lookup is done for each block (See Get KVC protocol below) starting at the last block and stopping when a match is found (alternatively we can use a local radix tree, see below)}
\item{If no match is found for a block the KVC for that block is split into fixed byte chunks}
\item{Each chunk is mapped into a separate LOS satellite using chunk\_id mod num\_los\_sats, the one with the fewest 
hops store chunk\_id 1}
\item{The mapping from server to chunk it is done with the current closest satellite as chunk 1. Then follows a pattern left to right top to bottom in concentric circles.}
\item{When a satellite is about to exit the LOS region all chunks stored need to be migrated to the satellite about to enter LOS in the same plane.}
\end{enumerate}

\begin{figure}[htbp]
	\centering
        \includegraphics[scale=0.5]{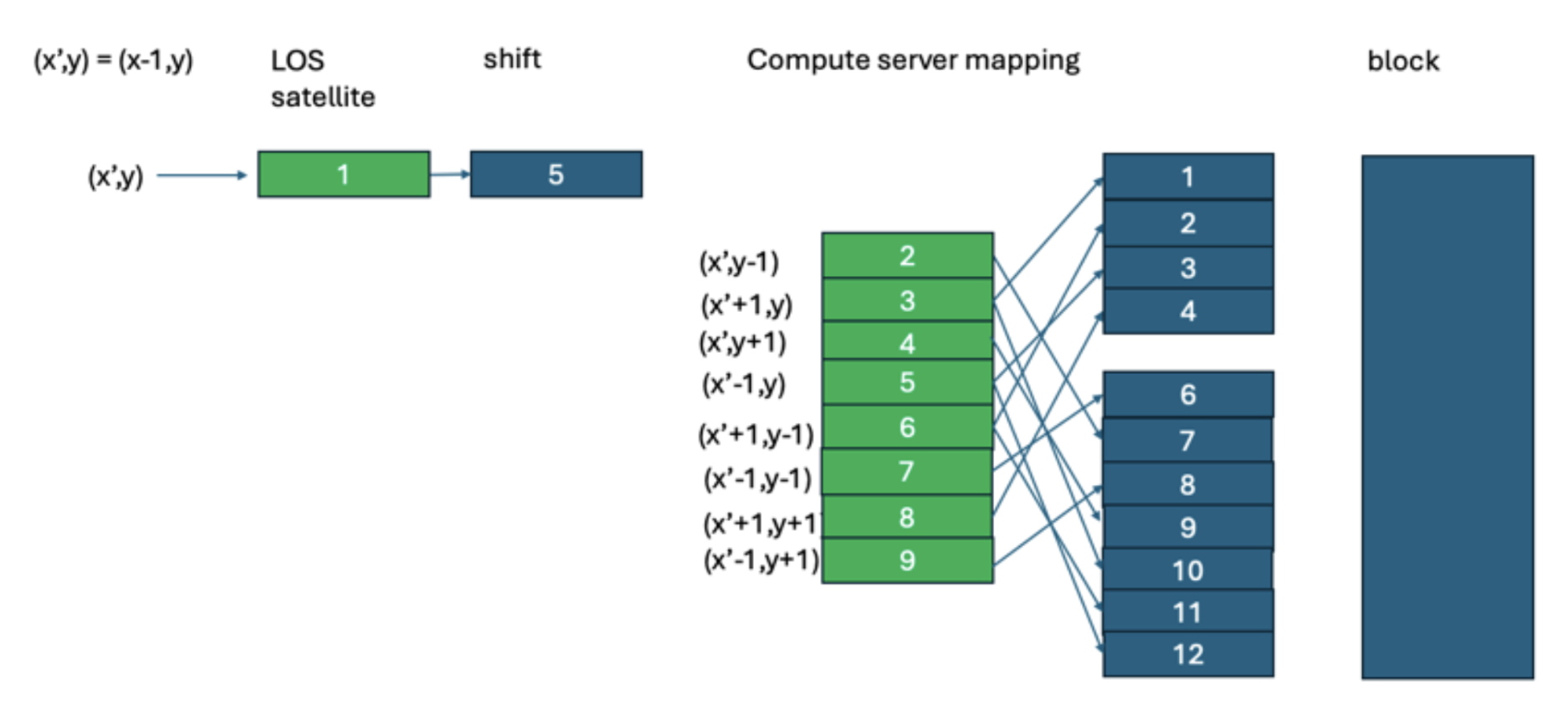}
        \caption{Getting Block After 1 Rotation: How a block can be recreated from chunks after one rotation. 
Note that rotations are predicable based on knowing the time of block creation.}
\label{getblock}
\end{figure}

\subsubsection*{Get KVC (Figure~\ref{getblock}):}
\begin{enumerate}
\item{See Step 1 in Set KVC}
\item{See Step 2 in Set KVC}
\item{The block list of hashes is searched with a binary search for a hash with chunk 1 within the satellite in LOS with fewest hops (or radix tree)}
\item{If a match is found the higher ordered half is searched}
\item{If a match is not found the lower ordered half is searched}
\item{If the higher ordered half is a single hash and it is not found the search stops with the result that the hash/block was not found and an empty KVC is used in the generation}
\item{The latest (highest ordered block) match is retrieved and all the chunks for that block are retrieved to reconstruct the KVC to be used in the generation}
\item{The lookups always start at the nearest satellite. It will return its chunk id and based on that the shift from left to right in the chunk to server mapping is found, and the server for all other chunks can be a computer and all chunks can be queried in parallel.}
\end{enumerate}

\subsection{Chunk Eviction}
When there is memory pressure, the LRU chunk will be evicted to make place for new block chunks. As soon as one
chunk is gone, the block it belongs to cannot be retrieved and must be purged. Hence an eviction needs to be
propagated to all the satellites holding the same chunks or at least the orbital plane that is used for lookups.
If a block only has one chunk, this is not an issue, and it could be a reason to keep the chunk size large as a
tradeoff for parallelism in retrieval and storage. 
The advantage of the concentric circle of storage of chunks is that all the chunks impacted by eviction are in the
direct neighborhood of the chunk initially being evicted, hence a simple gossip broadcast in all directions is sufficient.
Alternatively, lazy eviction can be implemented, where the lookup client will issue evictions when chunks in a block are
discovered to be missing. 
Yet another policy is to do cleanup of chunks that are not complete periodically. When chunks migrate, they can be evicted
so there is a natural eviction as part of the rotation synchronization as well.  

\subsection{Local Radix Block Index}
\begin{figure}[htbp]
	\centering
        \includegraphics[scale=0.5]{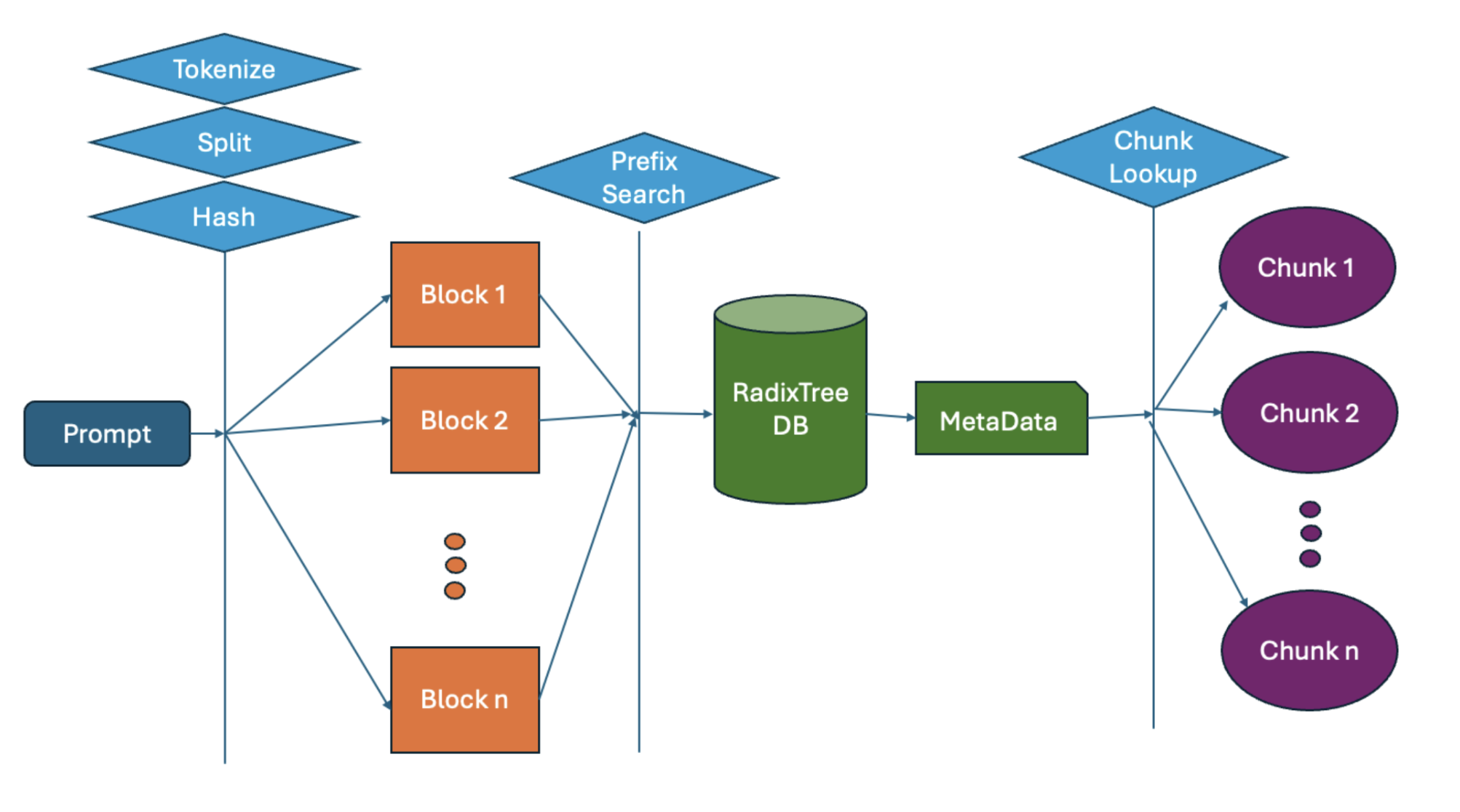}
        \caption{End-To-End process from a prompt to a chunk lookup using a radix tree.}
\label{radix}
\end{figure}

In the case of transformer KVC caching the ordered blocks resulting from the prompt need to be queried using a
longest prefix lookup. The last block in the block list matching will have the cache we are interested in. The lookup
may be done sequentially or with some binary search, but as an optimization we propose storing the block keys (not the
values) in a Radix Tree (a.k.a radix trie or prefix tree) locally were the LLM resides (on the ground or on-board the
satellite). An efficient lookup can thus be made in this local data structure to find out whether the block exist. The database entry for the block can also store meta data such as total number of chunks and the time of setting the value which would allow
the caller to compute where each chunk is currently located without having to query any external satellites. Eviction can
either be done when a local lookup succeeds but the values are not present in the constellation or by propagating
eviction broadcasts back to the LLM. Figure~\ref{radix} shows the full end-to-end process. We note that the radix tree lookup is an optional optimization,
and it is the only part that is specific to the Transformer (LLM KVC) application. All the other parts of the protocol
can be used as a general-purpose in-memory,Key Value Store (KVS) or distributed HashTable for any data where the key
can be converted to a string and the value is an arbitrary, potentially large, byte sequence.

\section{Simulation}

To validate our three server-to-satellite mapping approaches, we implemented a simulator that computes the worst-case latency based on the distance equation~\ref{eq:dm}, and the chunk farthest away. In the LOS case, all satellites could potentially be contacted in parallel, so the worst-case latency is a good indicator of the total time to both get and set chunks.

First, we need to establish some notation: suppose we want to route a chunk from the current satellite $s$ in the current orbital plane $o$ to the target satellite $s_t$ in the target orbital plane $o_t$.  First, we define the following distances from $(s,o)$ to $(s_t,o_t)$ along the north, south, west, and east directions:
% North Distance
\[
	d_{\text{north}}(o, o_t) =
\begin{cases}
o - o_t & \text{if } o_t < o \\
o + M - o_t & \text{if } o_t > o \\
0 & \text{otherwise}
\end{cases}
\]

% South Distance
\[
	d_{\text{south}}(o, o_t) =
\begin{cases}
o_t - o & \text{if } o_t > o \\
M - o + o_t & \text{if } o_t < o \\
0 & \text{otherwise}
\end{cases}
\]

% West Distance
\[
	d_{\text{west}}(s, s_t) =
\begin{cases}
s - s_t & \text{if } s_t < s \\
s + N - s_t & \text{if } s_t > s \\
0 & \text{otherwise}
\end{cases}
\]

% East Distance
\[
	d_{\text{east}}(s, s_t) =
\begin{cases}
s_t - s & \text{if } s_t > s \\
N - s + s_t & \text{if } s_t < s \\
0 & \text{otherwise}
\end{cases}
\]
The path from $(s,o)$ to $(s_t,o_t)$ begins with its chunk being routed to the neighbor of $(s,o)$ given by $(s+\Delta s, o+\Delta o)$, where
\[
(\Delta s,\Delta o) =
\begin{cases}
	(0, -1) & \text{if } d_{\text{north}}(o, o_t) < d_{\text{south}}(o, o_t) \\
	(0, 1) & \text{if } d_{\text{south}}(o, o_t) < d_{\text{north}}(o, o_t) \\
	(-1, 0) & \text{if } d_{\text{west}}(s, s_t) < d_{\text{east}}(s, s_t) \\
	(1, 0) & \text{if } d_{\text{east}}(s, s_t) < d_{\text{west}}(s, s_t) \\
(0, 0) & \text{otherwise.}
\end{cases}
\]
 
\begin{figure}[htbp]
        \centering
        \includegraphics[scale=0.3]{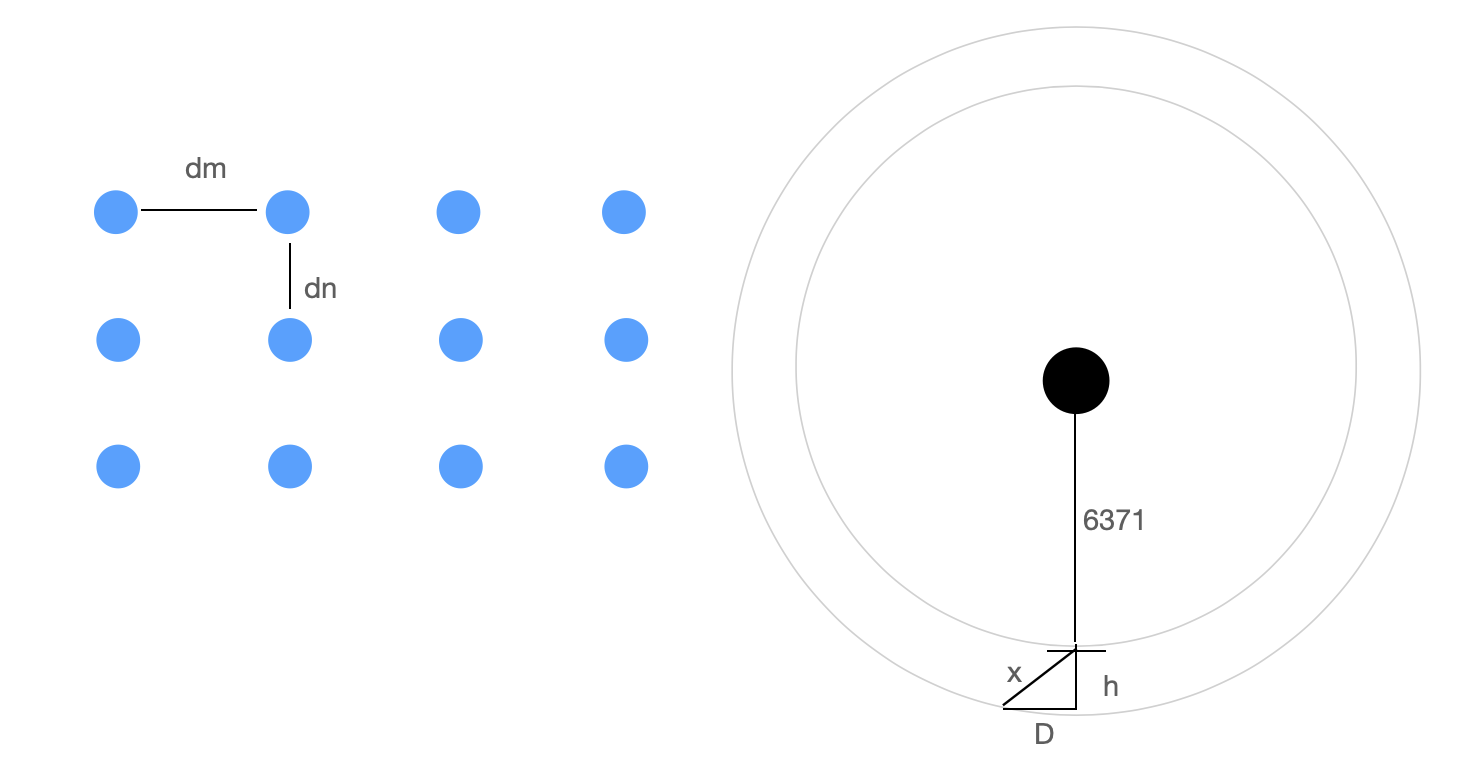}
        \caption{Simulated LOS latency geometry.}
\label{los}
\end{figure}

The shortest distance from one satellite to another is:
\begin{equation}
D = \sqrt{(D_m \Delta o)^2+(D_n \Delta s)^2}
\end{equation}
The distance from an LLM on earth to a LOS satellite can then computed as follows:
\begin{equation}
x = \sqrt{D^2+h^2}
\end{equation}
See Figure~\ref{los} for an illustration
of the parameters.

Table~\ref{T:simconfig} summarizes the simulation configurations.

\begin{table}[htbp]
        \caption{Simulation Configuration.}
\begin{center}
\begin{tabular}{|l|l|}
\hline
  Parameter & Values \\
\hline
  KVC\_BYTES & $2..21$ MB \\
  SERVERS & $9..81$ \\
  CHUNK\_PROCESSING\_TIME & $0.002..0.02$s \\ 
  ALTITUDE & $160..2000$km \\
  MAX\_SATELLITES & $15$ \\
  MAX\_ORBS & $15$ \\
  CENTER\_SATELLITE & $8$ \\
  CENTER\_ORB & $8$ \\
\hline
\end{tabular}
\label{T:simconfig}
\end{center}
\end{table}

\begin{figure}[htbp]
 \hspace{-2.1cm}
    \includegraphics[scale=0.2]{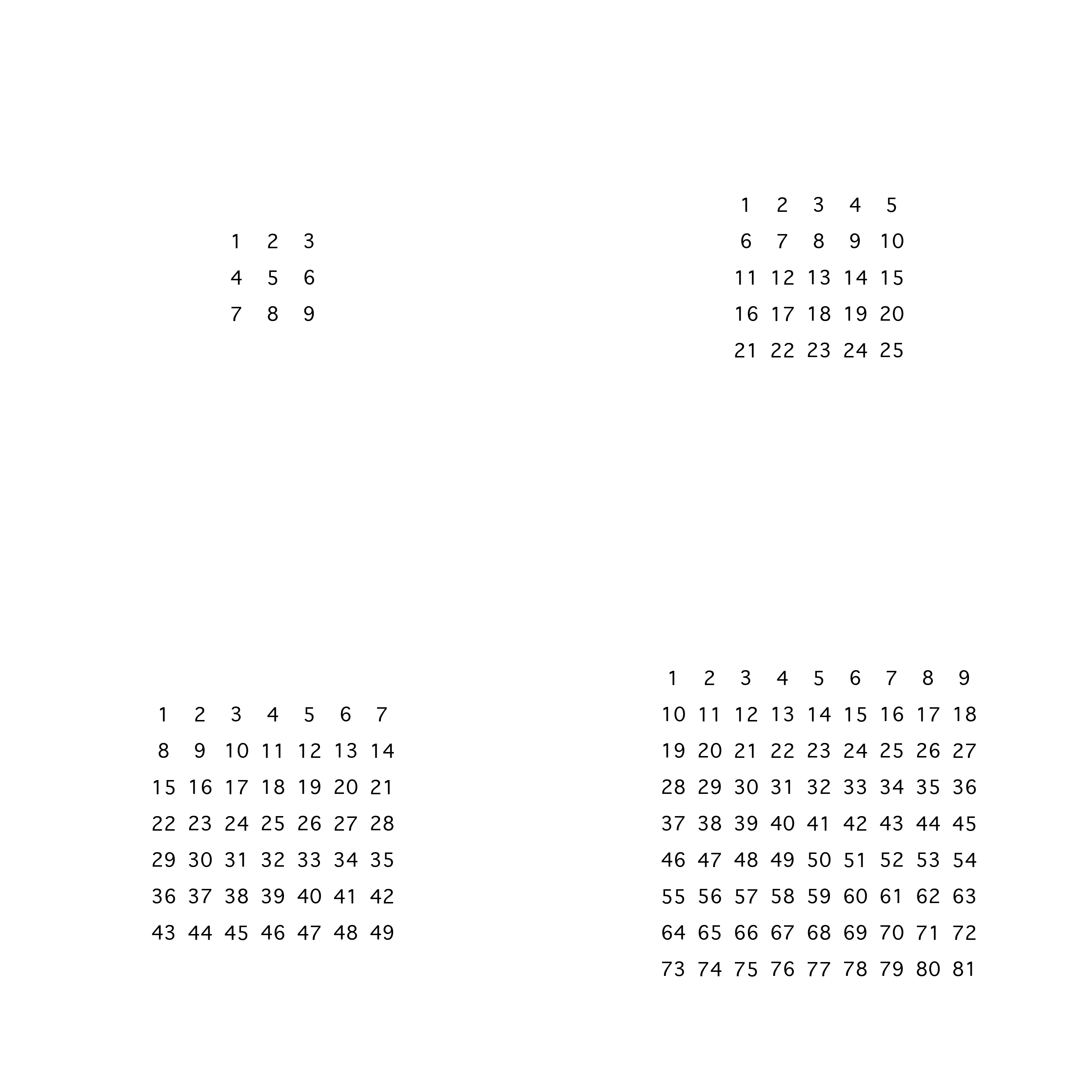}
        \caption{Rotation Aware Caching Mapping 3x3, 5x5 7x7, 9x9.}
\label{vizrotation}
\end{figure}

\begin{figure}[htbp]
        \hspace{-2.1cm}
        \includegraphics[scale=0.2]{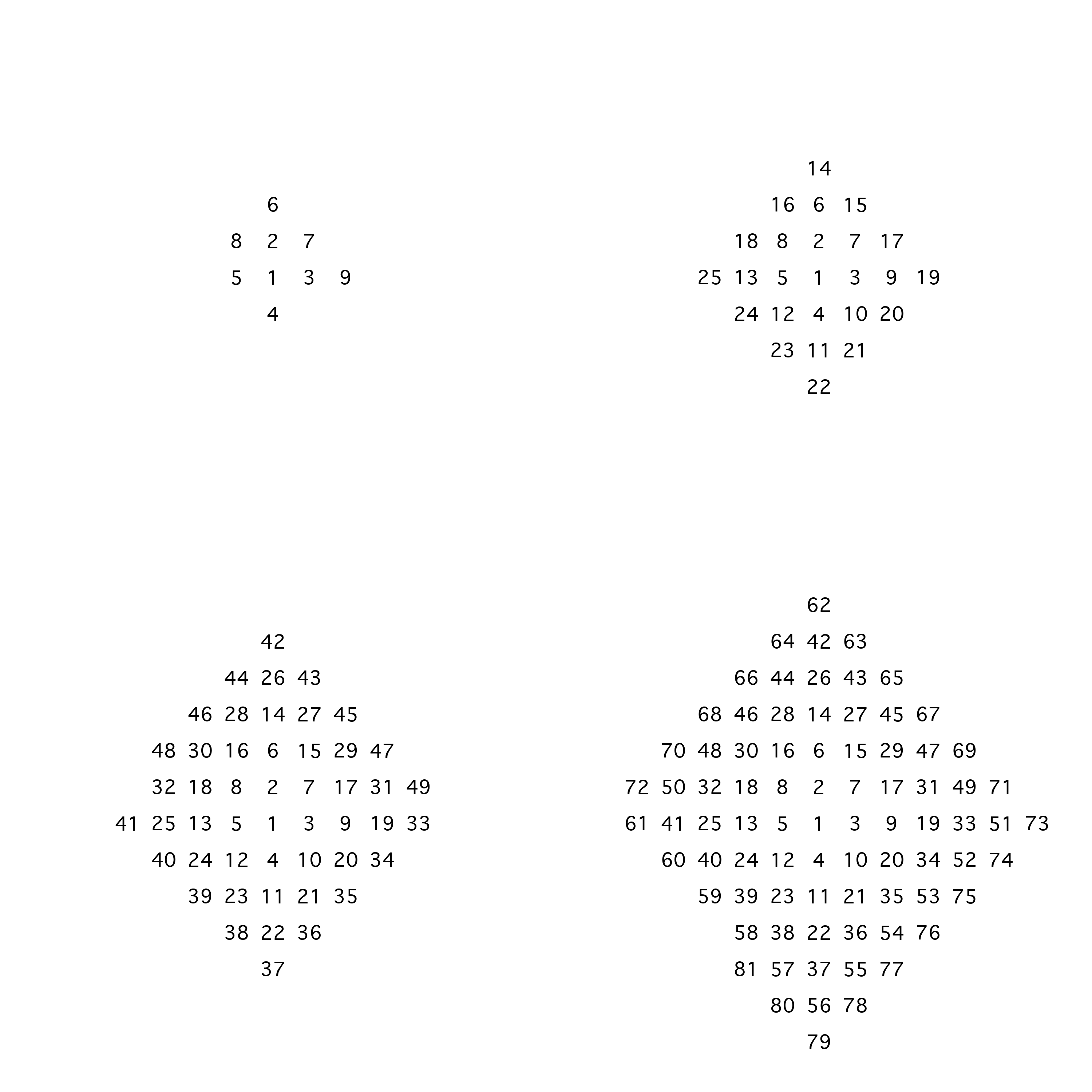}
        \caption{Hop Aware Caching Mapping 3x3, 5x5 7x7, 9x9.}
\label{vizhop}
\end{figure}

\begin{figure}[htbp]
       \hspace{-2.1cm}
        \includegraphics[scale=0.2]{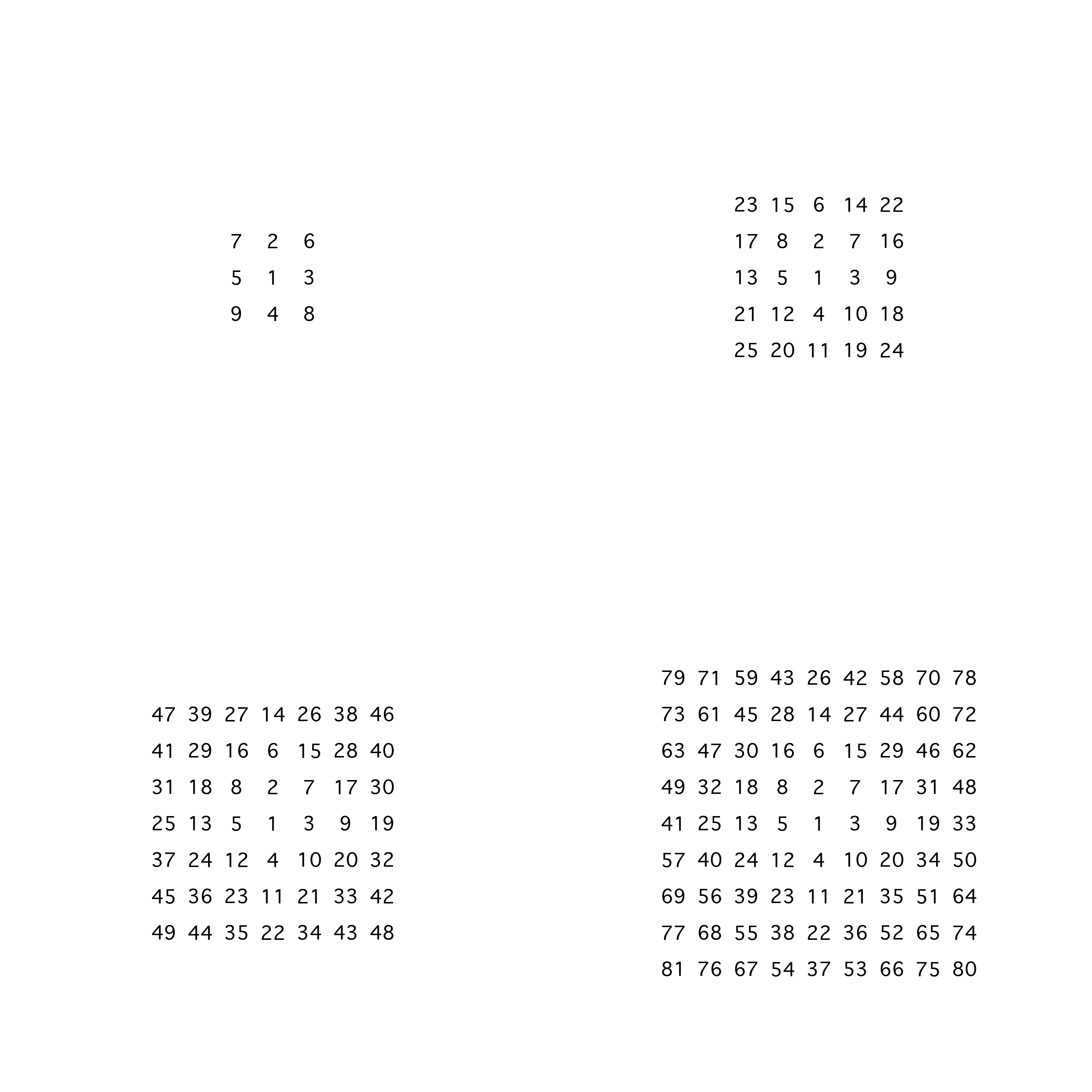}
        \caption{Rotation and Hop Aware Caching Mapping 3x3, 5x5 7x7, 9x9.}
\label{vizhoprotation}
\end{figure}

\begin{figure*}[t!]
        \subfloat[Altitude]{%
            \includegraphics[width=.48\linewidth]{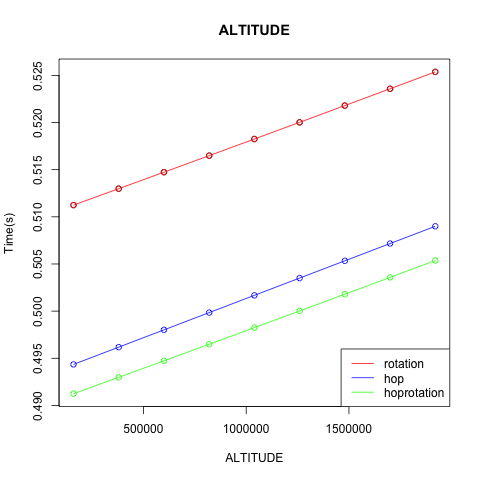}%
            \label{subfig:altitude}%
        }\hfill
        \subfloat[KVC Size]{%
            \includegraphics[width=.48\linewidth]{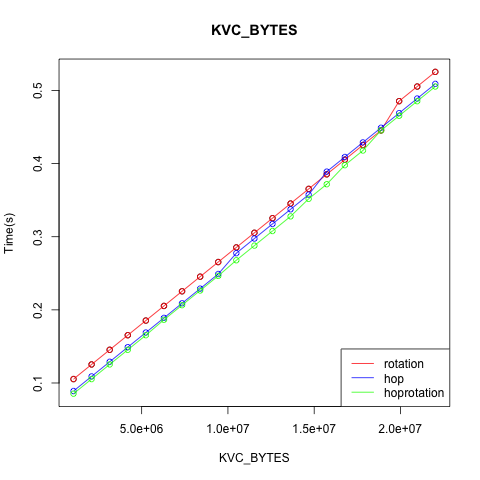}%
            \label{subfig:kvcsize}%
        }\\
        \subfloat[Servers]{%
            \includegraphics[width=.48\linewidth]{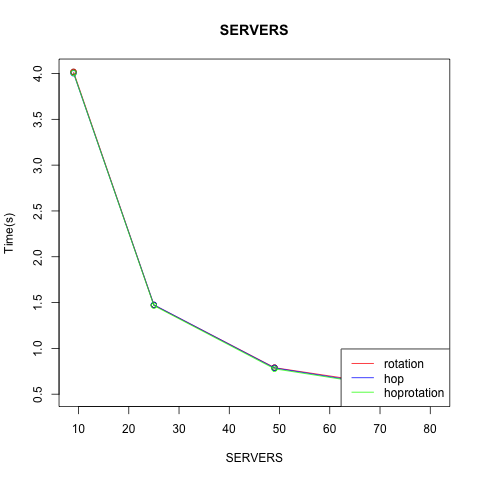}%
            \label{subfig:servers}%
        }\hfill
        \subfloat[Processing Time]{%
            \includegraphics[width=.48\linewidth]{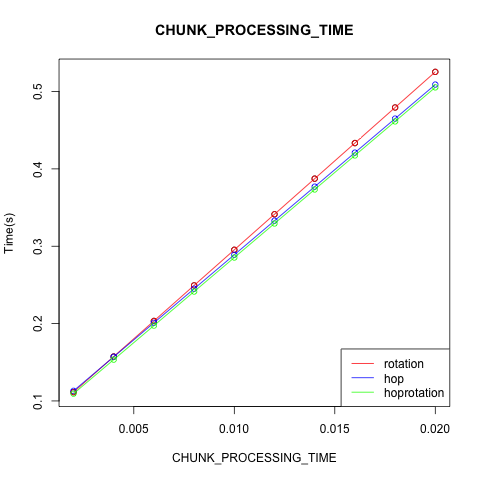}%
            \label{subfig:processingtime}%
        }
        \caption{Max Latency Across Parameters and Strategies}
        \label{fig:sim}
\end{figure*}

In Figure~\ref{fig:sim} we see that the hop- and rotation-aware approach results in lower latency than the hop-aware and the rotation-aware approaches across different altitudes. We also note that adding more servers reduces latency for all approaches. An 8x increase in servers results in about 90\% reduction in latency. For different KVC sizes and processing times the different strategies produce similar results. In summary, the simulations show that reducing the hops and adding servers are effective approaches to reducing latency with our KVC layouts.
\section{Implementation}
We have implemented the KVC protocol for a testbed comprising 5 Intel NUC Linux mini PCs hosting a 19x5 constellation, and an NVIDIA Jetson Nano 8GB GPU hosting the LLM.
The constellation is implemented as a NASA Core Flight System (cFS) deployment
with an added hashtable implementation and ISL routing from~\cite{sandholm2024,sandholm2025}. The communication between the LLM and the constellation is done over the CCSDS Space Packet Protocol over UDP~\cite{spacepacket2003}.

As a validation test we used the vLLM prefix caching benchmark example\footnote{\url{https://github.com/vllm-project/vllm/blob/main/benchmarks/benchmark_prefix_caching.py}}.
We hosted a 1B parameter LLM (TinyLlama/TinyLlama-1.1B-Chat-v1.0) on the GPU and fed it a ~250-character prompt as context. It resulted in 4x 128 token blocks of about 2.9MB each (with optimum-quanto 8bit quantization). The blocks were split into 6k chunks before storage and retrieval. We used 10 LOS cFS satellites to stripe the chunks across. This resulted in a 30 token generation speedup from 6.2s to 4.9s or 21\% when running the generation without and with the cache respectively. For a HQQ quantizer the speedup was about 24\%.

\begin{table}[htbp]
        \caption{Jetson cFS Testbed Experiment.}
\begin{center}
\begin{tabular}{|l|l|l|}
\hline
  Quantization & No KVC (seconds) & KVC (seconds) \\
\hline
  Optimum Quanto & $6.2$ & $4.9$ \\
  HQQ & $10.2$ & $7.8$ \\
\hline
\end{tabular}
\label{T:experiment}
\end{center}
\end{table}
\section{Conclusions}
We demonstrate the potential of leveraging the LEO edge as a caching layer to accelerate inference in transformer-based architectures, supported by both simulations and a proof-of-concept implementation. Our approach is applicable to KVCs in general, particularly in scenarios where KVC data is large, satellite memory is constrained, or communication bandwidth (both uplink and downlink) is limited. More broadly, our technique offers advantages for inference in any network with a dynamic topology between edge resources and clients. For instance, in edge computing in mobile wireless networks, it can proactively support seamless, low-latency user experiences by migrating KVCs as users (the clients) move between cells. Overall, our approach evaluates the feasibility and benefits of migrating the cache to the request, rather than requiring the request to find the cache.

\bibliographystyle{plain}
\bibliography{related}

\begin{thebibliography}{1}

\bibitem{spacepacket2003}
{Space Packet Protocol}; {Recommendation for Space Data System Standards (Blue
  Book)}.
\newblock Standard, {Consultative Committee for Space Data Systems (CCSDS)},
  National Aeronautics and Space Administration: Washington, DC, USA, 2020.

\bibitem{kwon2023}
Woosuk Kwon, Zhuohan Li, Siyuan Zhuang, Ying Sheng, Lianmin Zheng, Cody~Hao Yu,
  Joseph Gonzalez, Hao Zhang, and Ion Stoica.
\newblock Efficient memory management for large language model serving with
  pagedattention.
\newblock In {\em Proceedings of the 29th Symposium on Operating Systems
  Principles}, pages 611--626, 2023.

\bibitem{perdigues2021}
Josep Perdigues, Harald Hauschildt, Wael El-Dali, Silvia Mezzasoma, Monica
  Politano, Zoran Sodnik, and Christopher Vasko.
\newblock Hydron: The esa initiative towards optical networking in space.
\newblock In {\em 2021 European Conference on Optical Communication (ECOC)},
  pages 1--4. IEEE, 2021.

\bibitem{pfandzelter2022}
Tobias Pfandzelter and David Bermbach.
\newblock {QoS-aware resource placement for LEO satellite edge computing}.
\newblock In {\em {2022 IEEE 6th International Conference on Fog and Edge
  Computing (ICFEC)}}, pages 66--72. IEEE, 2022.

\bibitem{sandholm2025}
Thomas Sandholm, Sayandev Mukherjee, John Feland, and Bernardo~A Huberman.
\newblock {LeoDist: A Distributed Ledger On-board LEO Satellites}.
\newblock In {\em {2024 IEEE Space Computing Conference (SCC)}}, 2025.

\bibitem{sandholm2024}
Thomas Sandholm, Sayandev Mukherjee, and Bernardo~A Huberman.
\newblock A cloud in the sky: Geo-aware on-board data services for leo
  satellites.
\newblock {\em arXiv preprint arXiv:2410.07586}, 2024.

\bibitem{vanelli2024}
Alessandro Vanelli-Coralli, Nicolas Chuberre, Gino Masini, Alessandro Guidotti,
  and Mohamed~El Jaafari.
\newblock {\em 5G Non-{Terrestrial} Networks: Technologies, Standards, and
  System Design}.
\newblock Wiley Online Books, 2024.

\bibitem{vaswani2017}
Ashish Vaswani, Noam Shazeer, Niki Parmar, Jakob Uszkoreit, Llion Jones,
  Aidan~N Gomez, {\L}ukasz Kaiser, and Illia Polosukhin.
\newblock Attention is all you need.
\newblock {\em Advances in neural information processing systems}, 30, 2017.

\end{thebibliography}

\end{document}